\documentstyle[12pt,epsf,aps]{revtex} 
\normalsize

\begin{document} 
 
\title{\large {\bf Particles in classically forbidden area, 
neutron skin and halo, and pure neutron matter in Ca isotopes 
\thanks{This 
paper is dedicated to the memory of Prof. J. M. Hu who did 
so much to nurture the development of the nuclear physics 
in China. } 
} } 
\author{Soojae Im and J. Meng \\ 
     Department of Technical Physics, Peking University,\\ 
               Beijing 100871, P.R. China \\ 
      Center of Theoretical Nuclear Physics, National Laboratory of \\ 
       Heavy Ion Accelerator, Lanzhou 730000, China\\ 
         e-mail: meng@ihipms.ihip.pku.edu.cn} 
\date{\today} 
\maketitle 
 
\vspace{1.5cm} 
\begin{abstract} 
\indent 
The nucleon density distributions 
and the thickness of pure neutron matter  
in Ca isotopes were systematically studied 
using the Skyrme-Hartree-Fock model ( SHF ) from the 
$\beta$-stability line to the neutron drip-line. 
The pure neutron matter, related with the neutron skin or halo, was 
shown to depend not only on the Fermi levels of the  
neutrons but also on the orbital angular momentum of 
the valence neutrons. New definitions for the thickness  
of pure neutron matter are proposed. 
\end{abstract} 
 
{PACS numbers : 21.60.Jz, 21.65.+f, 21.10.-k, 21.10.Gv, 21.60.-n} 
 
Keywords: Neutron skin and halo, Skyrme-Hartree-Fock model, 
pure neutron matter, shell structure, Ca isotopes

 
The advent of radioactive ion beam(RIB) facilities provide us with new 
opportunities to create and study unstable nuclei far from the 
$\beta$ stability line.  They also enable us to reproduce the 
synthesis of elements that are normally only found in explosive supernova and 
neutron stars. This makes it possible to study the origin 
of the element abundance in the universe. As the number of unstable nuclei 
far exceeds the number of stable nuclei, experiments with radioactive beams 
were expected to open an unexplored region in the nuclear 
chart with promises of new phenomena and new symmetries 
quite different from those in the stable region. 
In particular, nuclei near the drip lines have valance 
nucleons that are very weakly bound and strongly coupled 
with the particle continuum \cite{MR96,NWD98}. The 
drip-line nuclei are expected to exhibit some exotic 
phenomena such as the vanishing of shell gaps and the occurence of neutron 
skins or halos. Some 
recent reports on the nuclear shell structure, radii, 
particle distribution, etc. in nuclei near the particle drip line  
can be seen in 
Refs.\cite{FOS92,SLHR94,DHNS94,WDN97,ME97,DNW96,Suz.98,NZFVHB98,Tan.85,MTY98} 
and the references therein. 
 
Here the main purpose is to address the implication of skins or 
halos such as: How does the shell structure influence the nuclear 
properties? To what extent does a pure neutron matter exist in 
a finite nucleus? How is the penetration of particle in the 
classically forbidden region related to the halo or 
skin? What is the role of the proton if pure neutron matter 
exists?  We will emphasize the general mechanism for the 
evolution of the nuclear size ( skin or halo) 
and the predication of 
pure neutron matter. The doubly magic $^{40}$Ca, $^{48}$Ca, and $^{60}$Ca 
nuclei were chosen for our calculations as they are good examples to 
illuminate our points.

 
The Skyrme-Hartree-Fock model ( SHF ) \cite{Sky.59,VB72,Rei.91},  
with some modifications (e.g.,  spin-orbit term ) \cite{RF.95}  
for exotic nuclei, is one of the  
most successful mean field approximations  
in Nuclear Physics. 
In this study the SHF calculations are performed with the force 
parameter set SIII\cite{BFGQ75} for the spherical case. 
The details of the calculation are similar to those found in Ref. \cite{Rei.91}. 
The computing code was written in C language, and checked with the 
results calculated by the Fortran code. 
For an open shell nucleus, a filling approximation has been used. 
To study nuclei near the drip line, the effects of    
pairing, continuum and deformation are very important; however, because we 
are focused on the general 
mechanism for the evolution of the nuclear size and the 
pure neutron matter, we are satisfied with results of the 
the SHF calculations.

 
We calculated the binding energy $E$ and two neutron separation 
energy $S_{2n}$ with the 
spherical SHF method for Ca isotopes. 
Compared with their experimental 
counterparts \cite{AW95}, 
the difference ($\Delta E$) between the  expeerimental and calculated 
binding energies is less then 3 MeV, i.e.,  
less than 1$\%$ of the experimental values ( $ \sim 400$ MeV ).

 
In Fig.1 we show the matter(a), proton(b) and neutron(c) densities 
for Ca isotopes. The same figures in logarithmic scale are given as 
inserts in order to show the exterior region of the density distribution. 
The exterior region lies outside of the potential barrier where  
it is classically forbidden for particle to enter. 
Because the exterior region is closely related with the skin or halo, 
studying the behavior of the density distribution in this region 
can give us more information on skins or halos. 
In the insert of Fig.1(a), we can see that the matter density  
in the exterior region 
increases with mass number. 
Because the proton density in the exterior region decreases with 
mass number, as shown in the insert of Fig.1(b), it is clear 
that the matter density in the exterior region is mainly determined by 
the neutron density in the neutron-rich nuclei,  
as seen in Fig.1(a) and Fig.1(c). 
We can also see an abrupt change between $^{48}$Ca and $^{54}$Ca as well as a 
decrease of mass density in the tail when going from 
$^{54}$Ca to $^{60}$Ca. 
It seems surprising that $^{54}$Ca has a longer tail than the 
isotopes with $A > 54$. This can be 
partly understood from the two neutron separation energy $S_{2n}$. 
The $S_{2n}$ for $^{54}$Ca is smaller than 
$^{60}$Ca, similar to the case of $^{6}$He and $^{8}$He. 
Further understanding can be obtained from the centrifugal barrier  
shown in Fig.4 and the discussion below.

The decrease of the proton density in the exterior region can be 
understood from the single-particle energy. 
In Fig.2(a) one can observe that the single-particle energies for protons 
decreases with 
mass number. 
The increase of binding energy for protons is due to the p-n 
interaction, which is stronger than the p-p interaction  
and increases as 
more neutrons are added. 
As results protons are more and more bound and  
it becomes more difficult to penetrate into the 
classically forbidden area. 
Therefore the proton density in the exterior region 
decreases. 
 
In Fig.2 (b), the single-particle energies for neutrons are plotted and  
the shell closure in the levels for Ca isotopes can be seen. 
The last valence neutrons in $^{40}$Ca, $^{48}$Ca and $^{60}$Ca are 
filled in the $2s1/2$, $1f7/2$, and $1f5/2$, respectively. 
The orbits $2s1/2$ and $1f7/2$ mark the neutron magic numbers $N = 20$ 
and $28$, respectively, thus $^{40}$Ca and $^{48}$Ca are closed shell 
nuclei. 
The $1f5/2$ orbits is relatively well separated from the 
$1g9/2$ orbital, but the energy gap between the $1f5/2$ and $1g9/2$  
is about 3 MeV. 
It is small compared to the 5 MeV separation between the $2s1/2$ and $1f7/2$, and 
the 7 MeV gap between the $1f7/2$ and $2p3/2$. 
So the $1f5/2$ orbital marks a weak shell closure or sub-shell closure 
corresponding to the neutron number $N = 40$.

Comparing with the proton case, the single-particle energies for  
neutrons remain the same with the mass number.  
So the increase of  the neutron density in the exterior region  
is not due to the change of single-particle energy. 
To understand the neutron density in the exterior region,  
We show the relative contributions of the different  
orbits to the neutron density for (a)$^{48}$Ca, (b)$^{54}$Ca  
and (c)$^{60}$Ca in Fig.3.  The corresponding figures in logarithmic scale  
are given as inserts. 
From Fig.3(a) we see that the neutron density in the 
exterior region for $^{48}$Ca is mainly determined by the 1f7/2. 
It has a higher density than the 2s1/2 which determines the neutron density 
in the exterior region for $^{40}$Ca, but has almost the same slope as 
the 2s1/2 in the exterior region,  even though they have quite different 
bindings. 
As a result, the neutron density of $^{48}$Ca doesn't change much in the 
exterior region compared to that of $^{40}$Ca, as shown in Fig.1(c). 
In contrast to this, the 2p orbits have higher  
densities in the exterior region compared 
with the 1f7/2 [see Fig.3(b)]. 
The 2p orbits play a main role in the 
exterior region for $^{54}$Ca, which is the reason for the abrupt change between $^{48}$Ca and 
$^{54}$Ca. 
In Fig.3(c), we can also see that the 1f5/2 contributes less to the density 
in the exterior region than the 2p orbits, even though it has less 
binding than the 2p orbits in $^{60}$Ca as seen in Fig.2(b). 
The contributions of the 2p orbits to the density in the exterior region 
are also hindered by their stronger bindings as shown in Fig.2(b). 
So the $^{60}$Ca nucleus has relatively lower density in the outer region beyond 
10 fm than $^{54}$Ca as shown in Fig.1(a) and (c).

Now we turn to the reason for the dominant contribution of the 2p 
orbits on the density in the exterior region. 
In Fig.4(a), the neutron potential curves for the $p$, $d$ and $f$ 
orbits in $^{48}$Ca are given. 
The potential barriers are also given as an insert in a different scale. 
From the figure we can see that the potential barrier of the $p$ orbits 
is lower than that of the $f$ orbits 
because of the small centrifugal potential. 
Although the potential barriers decrease with mass number  
in Fig.4(b), the difference of the potential barrier between the 
$f$ orbits and the $p$ orbits is still around 4.5 Mev for all 
isotopes. 
This means that particles in a smaller $l$ orbit will have more chance to 
tunnel than particles in a large $l$ orbit. 
 
From these results it is clearly shown that the potential barrier and 
level energy are main factors for determining the density distributions 
in the exterior region. 
Therefore even though particle separation energy is small, the 
corresponding nucleus is not necessary a halo nucleus if the last few 
particles are in large $l$ orbits.

 
Concerning the halo phenomena, the definition of a halo is one of 
the most interesting topics. 
Here we suggest to define the thickness of a halo 
using the density distribution in the exterior region quantitatively. 
The one-particle and zero-particle radii  
$R_{\rm N_q=1,0.1}$ are defined as:  
$\int_{R_{\rm N_q =1}}^\infty 4 \pi r^2 \rho_{p/n} ~dr = 1$, 
and 
$\int_{R_{\rm N_q=0.1}}^\infty 4 \pi r^2 \rho_{p/n} ~dr = 0.1$. 
Now the thickness of neutron matter $t_n$ can be measured by the radii 
defined above. 
The radius $R_{\rm Z=0.1}$ is considered as a position 
beyond which the number of protons become zero, then  
$t_n= R_{\rm N=0.1} - R_{\rm Z=0.1}$  
is the thickness of pure neutron matter. 
In Fig.5,  we show $t_n$ which is given  by  
[A] $R_{\rm N=0.1} - R_{\rm Z=0.1}$; 
[B] $R_{\rm N=1} - R_{\rm Z=0.1}$;  
[C] $R_{\rm rms,n}-R_{\rm rms,p})$, i.e., the difference of neutron and  
proton $rms$ radii.  
 
In curve [A], $t_n=R_{N=0.1}- R_{Z=0.1}$ shows a dramatic  
increase from 
$^{50}$Ca to $^{54}$Ca, and becomes a constant value after $^{54}$Ca. 
The dramatic increase from $^{50}$Ca to $^{54}$Ca is due to the 
contributions of the neutron 2p orbits to the density in the exterior 
region as discussed before. 
So $R_{N=0.1}$ and $t_n$ will increase dramatically. 
The constant value after $^{54}$Ca is due to the high centrifugal 
potential barrier of the neutrons 1f5/2 and 1g9/2 orbits which are 
filled after $^{54}$Ca. 
Because these orbits contribute less to the density in the exterior 
region than the 2p orbits, the density in the exterior region is 
determined mainly by the 2p orbits and remains unchanged  
after $^{54}$Ca.

In curve [B] $t_n=R_{N=1} - R_{Z=0.1}$ represents the area 
where no protons exist, but still one neutron exists. 
It increases gradually and becomes positive after $^{54}$Ca. 
The positive values imply a neutron skin or halo in the nuclei after 
$^{54}$Ca. 
Comparing to the curve [A], we can not find any sign of shell effects in 
curve [B].   
Compared with curve [A], the curve [C]$t_n=R_{\rm rms,n}-R_{\rm rms,p}$ 
does not show any shell effects as in curve[A] and furthermore  
the meaning of the sign change is not so clear as in curve [B]. 
 
From these results it is shown that $t_n=R_{N=0.1} - R_{Z=0.1}$ and 
$t_n=R_{N=1} - R_{Z=0.1}$ can be used as a quantity to measure the 
thickness of the neutron skin or halo. 
And the difference of neutron and proton $rms$ radii 
$t_n=R_{rms,n}-R_{rms,p}$ is not a preferable quantity to measure the 
thickness of pure neutron matter.


Summarizing our investigations,   
the ground state properties of Ca isotopes were systematically 
studied with SHF from $^{40}$Ca to $^{70}$Ca and a new  
definition for the thickness of skins or halos is  
proposed. We found that: 
a) The shell effects were clearly demonstrated in the 
particle density distributions and two neutron separation energy.
b) The shell effect played an important role in the neutron density 
distribution, particularly in the tail. 
c) The centrifugal potentials were essential at large distances. 
The potential barriers and binding energies will determine the density 
distributions in the exterior region. Halo nuclei must have a nucleon  
in a low $l$ orbitals such as $s$ and $p$. 
d) The thickness of pure neutron matter was defined with different 
reference positions for the proton and neutron zero particle radius. 
The definitions  $t_n=R_{N=0.1} - R_{Z=0.1}$ and 
$t_n=R_{N=1} - R_{Z=0.1}$ are better to define  
the thickness of the neutron skin or halo than  
the difference of neutron and proton $rms$ radii. 
We should say that the effects of pairing, continuum, 
deformation, etc. that were neglected in the present study,  
are very important and need to be treated carefully. The work 
along this line is currently being performed and will be 
addressed in our future works. 
 
The authors thank Peter Ring for useful discussion and 
commentsi,  Daryl Hartley and Yang Sun for their careful 
reading of the manuscript and helpful discussion. 
This work was partially sponsored by the National Science 
Foundation in China under Project No. 19847002,  by SRF 
for ROCS, SEM, China and by the Research Fund for the Doctoral 
Program of Higher Education (RFDP).


\newpage 
 
\begin{figure} 
\caption{ The matter(a), proton(b) and neutron(c) density distributions 
in Ca isotopes. Same figures in logarithmic scale are given as inserts.} 
\label{Density} 
\end{figure} 
 
\begin{figure} 
\caption{ (a)Proton single particle energies and (b)neutron single 
particle energies as a function of mass number in Ca isotopes.} 
\label{SPE} 
\end{figure}

\begin{figure} 
\caption{The contributions of each orbit to the neutron density as a 
function of radius in (a)$^{48}$Ca, (b)$^{54}$Ca and (c)$^{60}$Ca. Same 
figures in logarithmic scale are given as inserts.} 
\label{SPDens} 
\end{figure}

\begin{figure} 
\caption{ (a)The potentials for each neutron orbit in $^{48}$Ca. 
(b) The isospin dependence of potential barrier, i.e., the sum of mean 
field potential and centrifugal potential, for neutron p, d and f orbits 
in Ca isotopes.} 
\label{SPPot} 
\end{figure} 
 
\begin{figure} 
\caption{  
The thickness of neutron matter $t_n$ i in Ca  
isotopes measured by [A] $R_{N=0.1} - R_{Z=0.1}$ (solid circles);  
[B]$R_{N=1} - R_{Z=0.1}$ (open circles); 
[C]$R_{\rm rms,n}-R_{rms,p}$ (solid rectangles).} 
\label{NMThick} 
\end{figure}

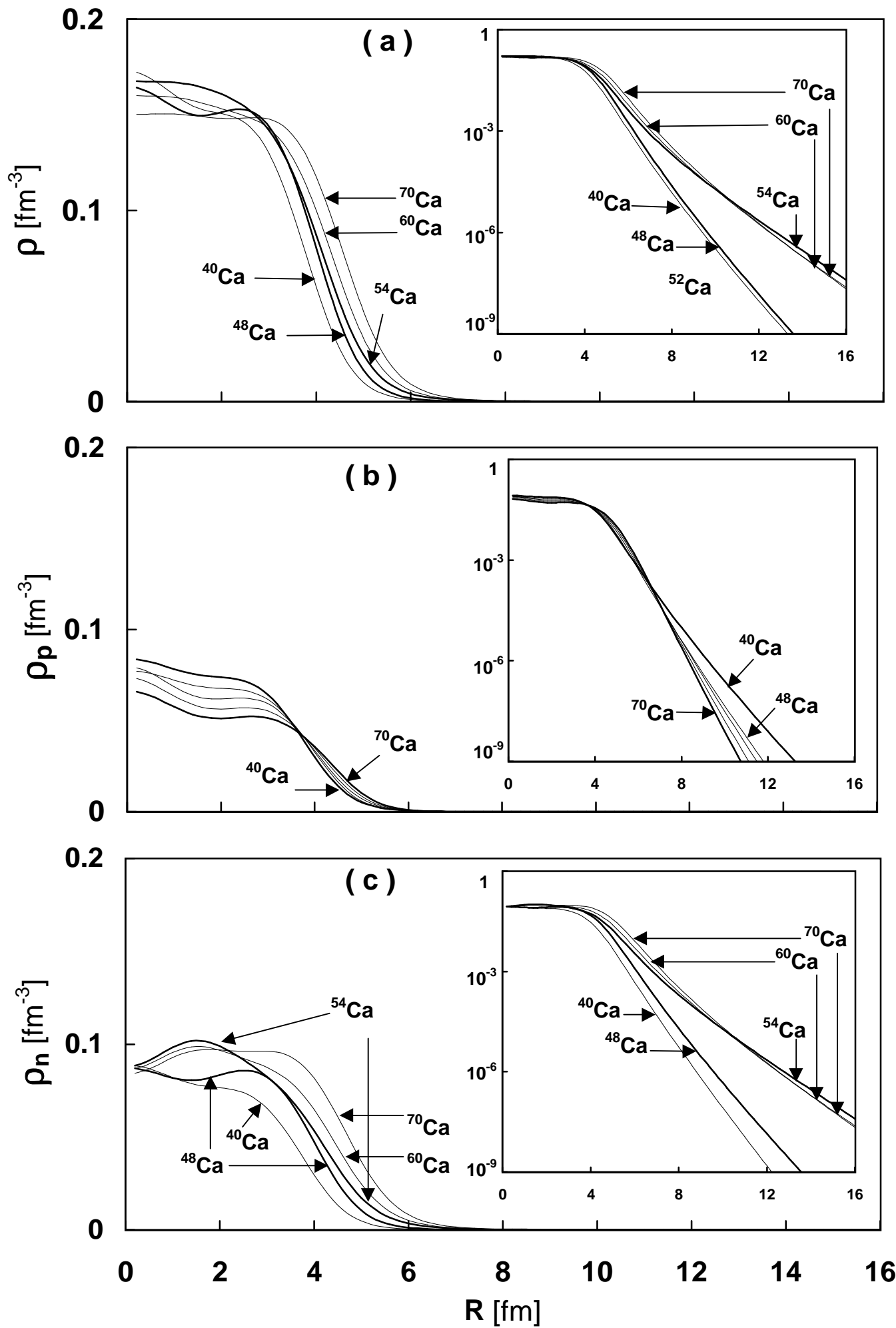
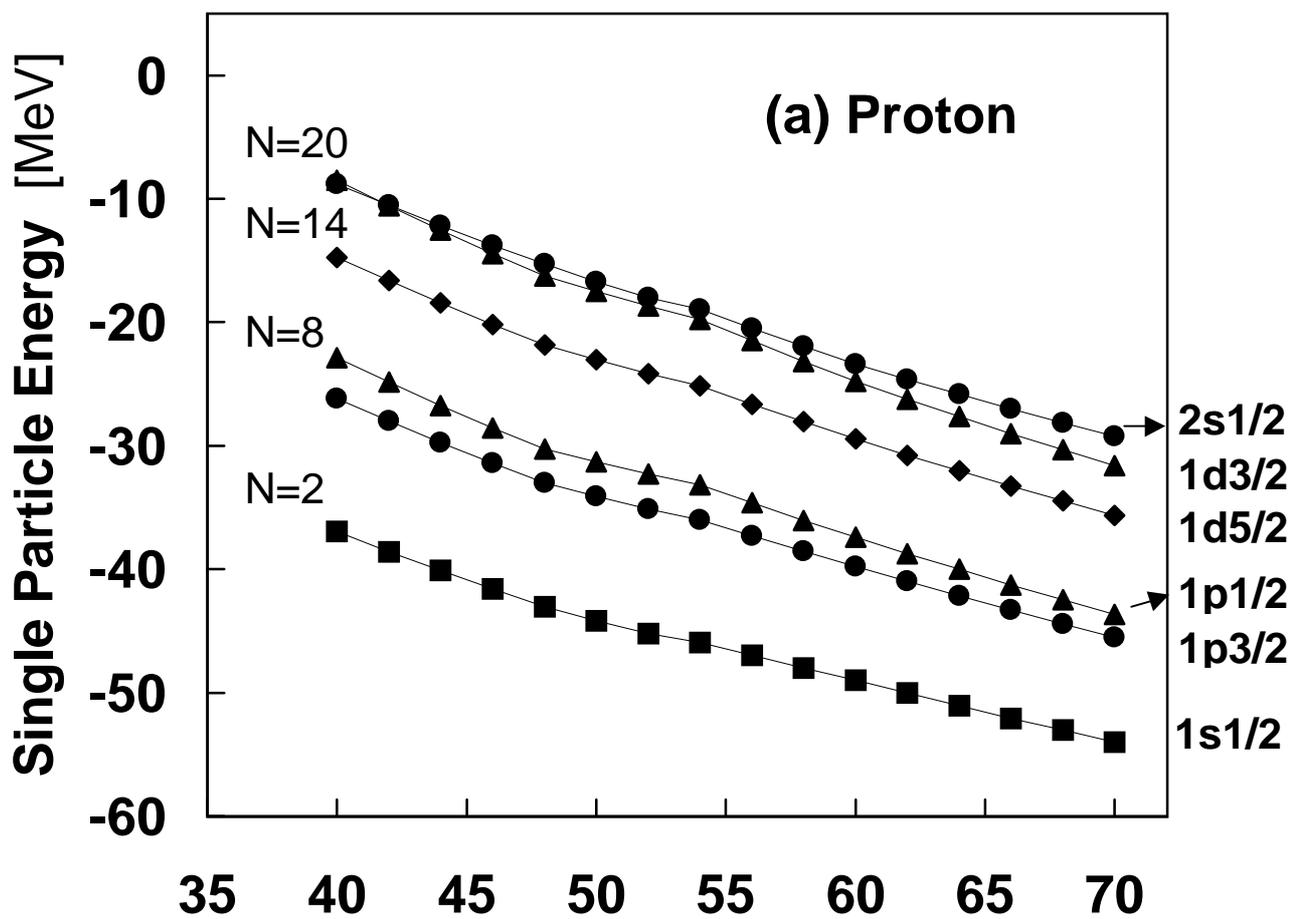
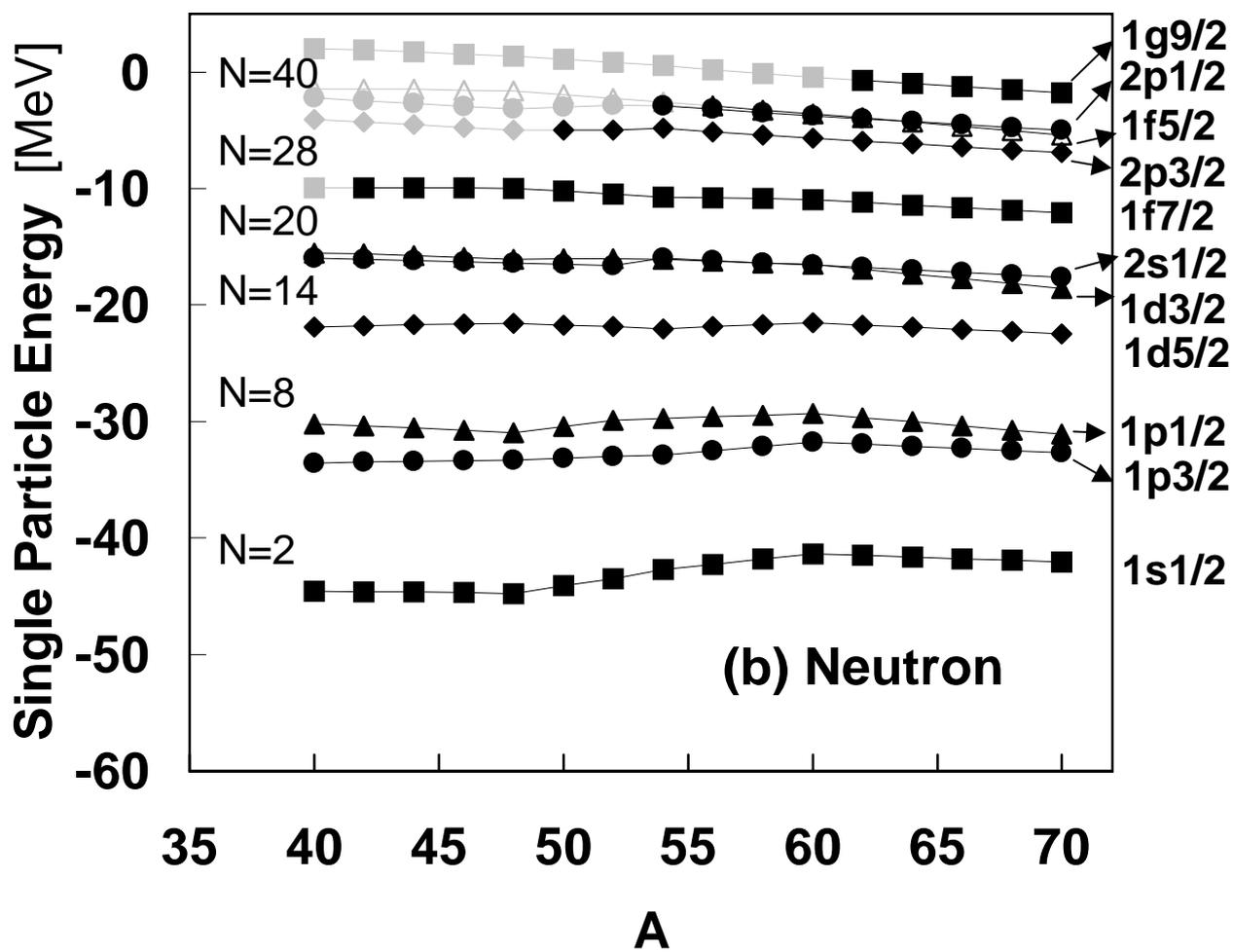
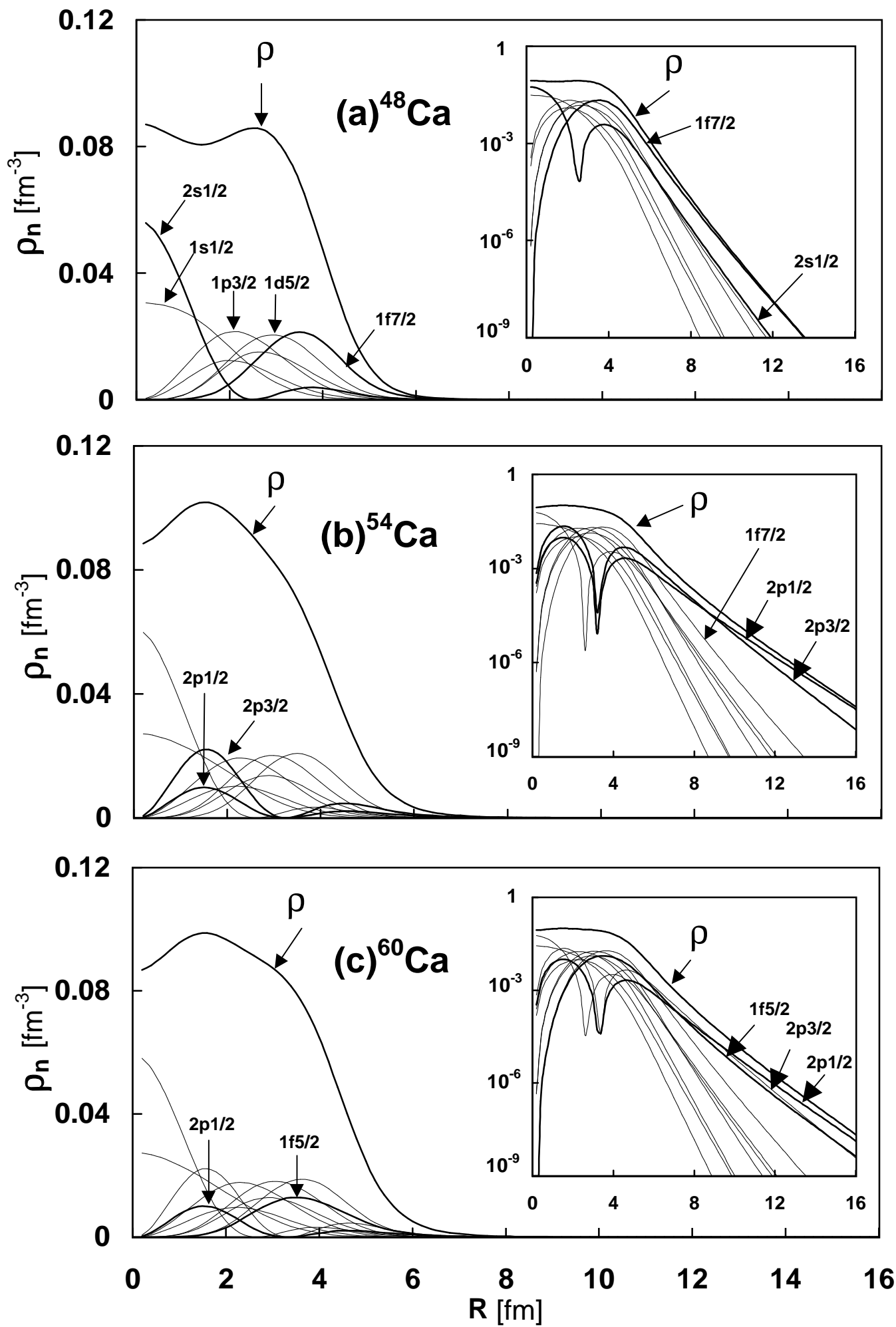
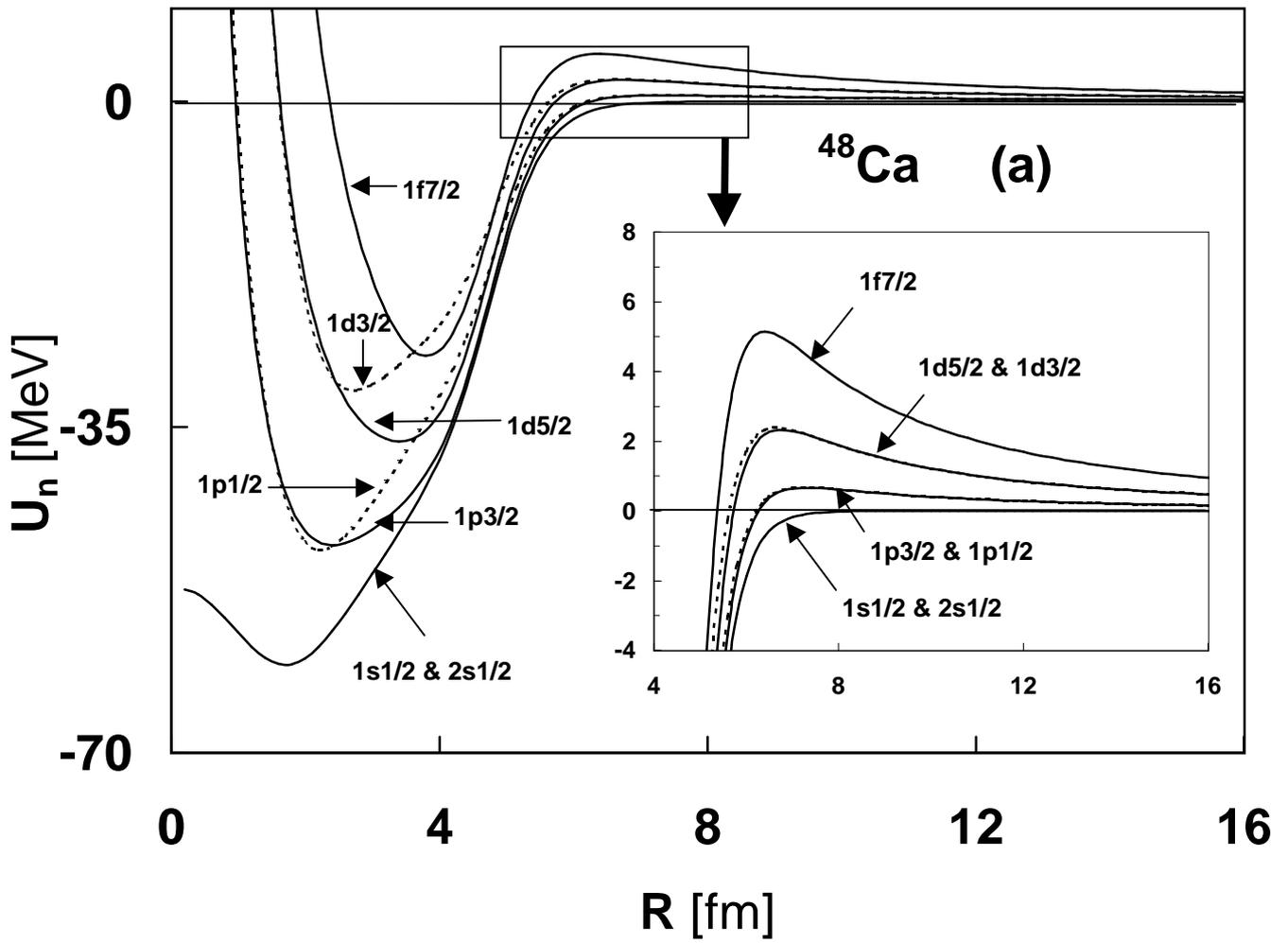
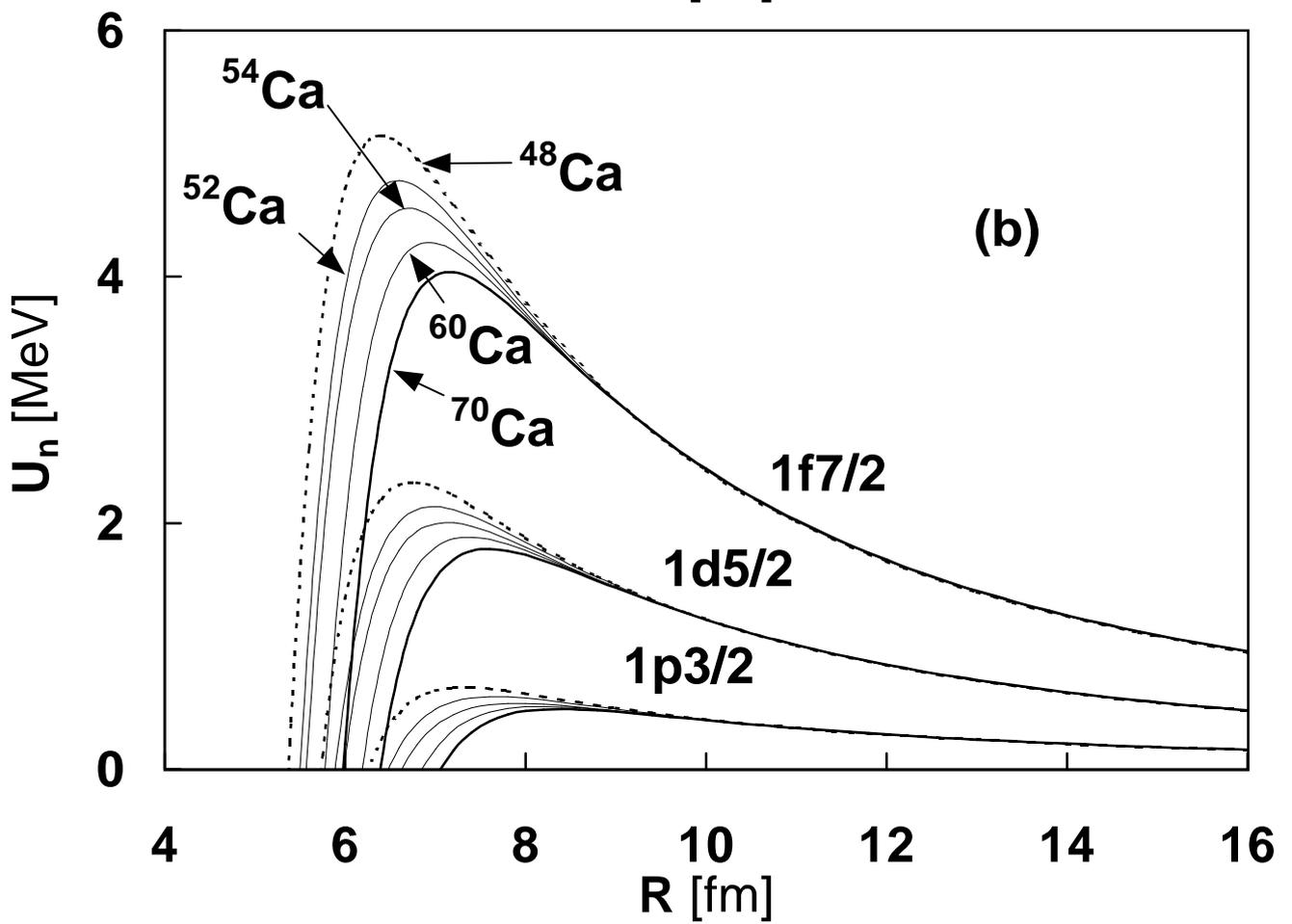
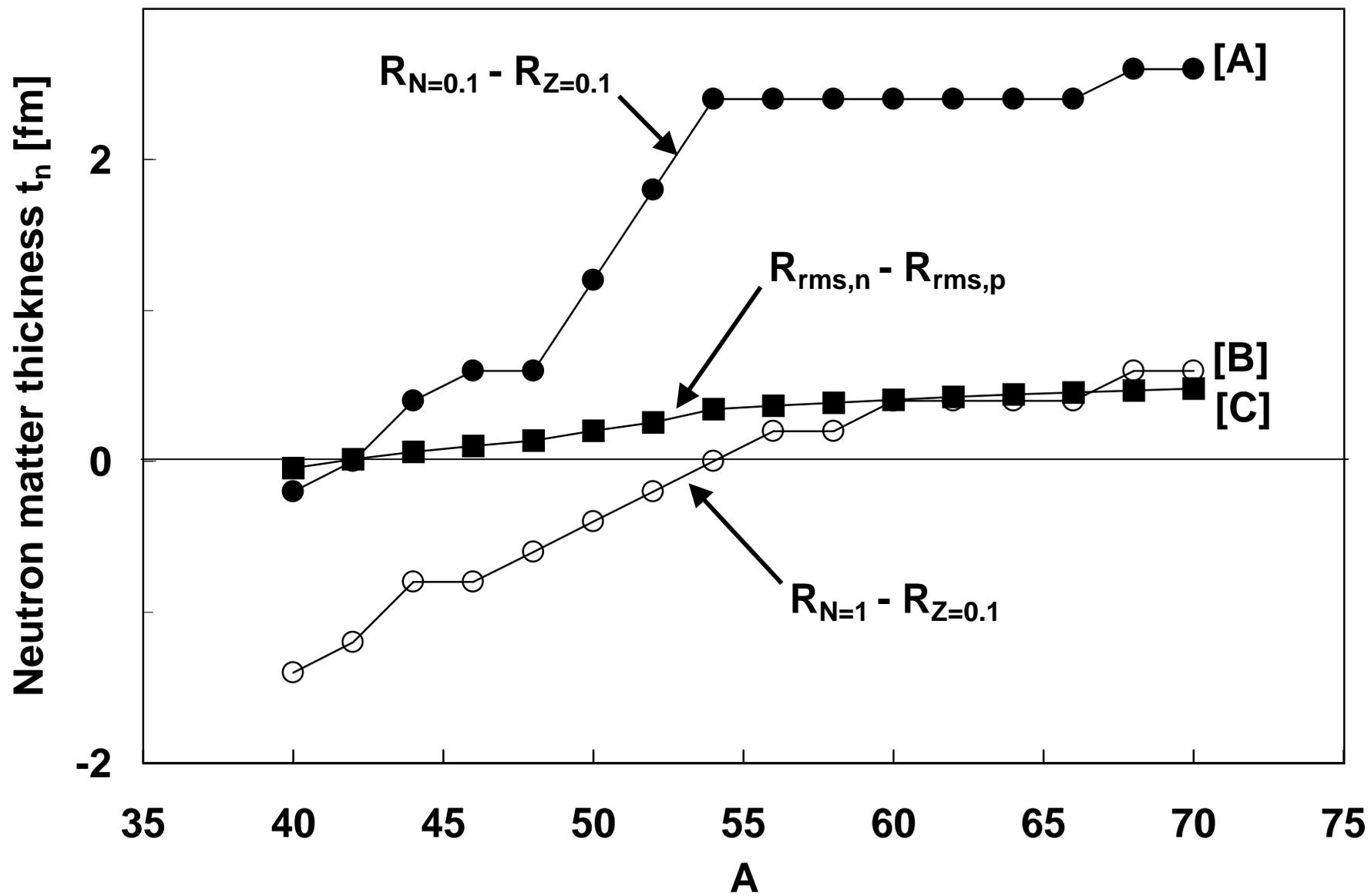
\end{document}